\documentclass[aps,prl,twocolumn,nofootinbib,
superscriptaddress,
]{revtex4-1}

\usepackage{hyperref}
\usepackage{url}
\usepackage{setspace}
\usepackage{amsmath,amssymb,amscd,braket}
\usepackage{amsfonts}
\usepackage{color}
\usepackage{graphicx}
\usepackage{xcolor}

\newcommand{\bea}{\begin{eqnarray}}
\newcommand{\eea}{\end{eqnarray}}
\newcommand{\be}{\begin{equation}}
\newcommand{\ee}{\end{equation}}
\newcommand{\ba}{\begin{align}}
\newcommand{\ea}{\end{align}}

\begin{document}
\title{
Price's law from quasinormal modes
}

\author{Paolo Arnaudo}
\email{p.arnaudo@soton.ac.uk}
\affiliation{Mathematical Sciences and STAG Research Centre, University of Southampton, Highfield, Southampton SO17 1BJ, UK}

\author{Benjamin Withers}
\email{b.s.withers@soton.ac.uk}
\affiliation{Mathematical Sciences and STAG Research Centre, University of Southampton, Highfield, Southampton SO17 1BJ, UK}

\begin{abstract}
We show that Price's power-law tail for perturbations of Schwarzschild, $t^{-2\ell-3}$ as $t\to \infty$, can be obtained from a sum of Schwarzschild-de Sitter quasinormal modes in the limit $\Lambda \to 0^+$. 
\end{abstract}

\maketitle

\section{Introduction and results}

The retarded Green's function for linear perturbations around the four-dimensional Schwarzschild black hole is well known to display a power-law decay at late times, which is known as Price's law \cite{Price1, Price2, DONNINGER2011484, Price:2004mm}. 
This behavior can be attributed to a branch cut of the Green’s function in the frequency domain.

Analytic investigations of the contribution of the branch cut in the frequency plane were carried out at large radius \cite{leaver1986, ching-PhysRevD.52.2118, Andersson:1996cm}, at large frequency via WKB techniques \cite{MaassenvandenBrink:2003as, Casals:2011aa}, or by using series of confluent hypergeometric functions evaluated on the negative imaginary axis \cite{casals1, Casals:2012ng, Casals:2015nja}.
The difficulty of analytically treating the branch cut contribution is due to the presence of an irregular singularity at radial infinity, around which there are no solutions admitting a power series expansion with a finite radius of convergence. 

\begin{figure}[t]
    \centering
    \includegraphics[width=\columnwidth]{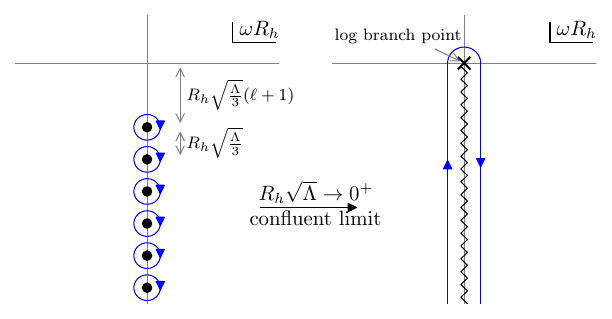}
    \caption{The analytic structure of the retarded Green's function in frequency space, $\widetilde{G}(\omega, r, r')$, at small $|\omega R_h|$, illustrating the change caused by heating up null infinity with a small $\Lambda > 0$. \textbf{Left panel:} Schwarzschild-de Sitter at small but finite $\Lambda >0$, which contains QNM poles at de Sitter frequencies \eqref{dSmodes} (black points), as well as black hole QNMs at order-1 values of $\omega R_h$ (not shown). \textbf{Right panel:} Schwarzschild, showing a logarithmic branch cut obtained in the limit $\Lambda \to 0^+$, from the coalescence of poles in the left panel. We demonstrate that the residue sum (blue contour, left panel) reproduces the power-law tail, $t^{-2\ell-3}$, equal to the branch cut discontinuity integral in this limit (blue contour, right panel).}
    \label{fig:analytic_structure}
\end{figure}

In this work, we overcome this issue by considering the same class of linear perturbations around the four-dimensional Schwarzschild-de Sitter solution with a small cosmological constant $\Lambda$. In this way, all singular points of the perturbation equations are regular, and there always exist local solutions with finite radius of convergence. Moreover, the branch cut structure present in the Green's function in the frequency plane is resolved in a set of poles along the imaginary axis, whose spacing is proportional to $\sqrt{\Lambda}$. See FIG. \ref{fig:analytic_structure} for illustration.
The computation of the branch cut contribution is then reduced to a sum of residues of these poles contributions, that, by taking the $\Lambda\to 0^+$ limit, matches the expected time dependence of Price's law, and with the correct coefficient.

Mathematically, the situation is analogous to charged scalar perturbations of the low-temperature Reissner-N\"ordstrom-AdS$_5$ black hole, a mechanism we previously detailed in \cite{Arnaudo:2024sen}. In the present work, temperature is introduced by turning on a small $\sqrt{\Lambda}$, which serves to heat up null infinity, converting the irregular singular point there into a regular one.\footnote{Both are governed by a confluence limit of the Heun equation, and moreover there is a direct mapping between the perturbation equations for a small black hole in AdS$_2$, and those appearing in the far region of Schwarzschild-de Sitter at small $\sqrt{\Lambda}$.}

The basic analytical mechanism is most easily illustrated by turning to a computation in the near-horizon region. At small $\Lambda >0$ the frequency-space Green's function, $\widetilde{G}(\omega, r, r')$ contains a set of quasinormal mode (QNM) poles at frequencies
\begin{equation}
\omega_n^{(\text{dS})}=-i\,\sqrt{\frac{\Lambda}{3}}\left(\ell+n+1\right)+ O(\Lambda), \quad n\in \mathbb{Z}_{\geq 0},\label{dSmodes}
\end{equation} 
which we refer to as the de Sitter QNMs.
See the left panel of FIG. \ref{fig:analytic_structure}.
Each such mode has a residue in $\widetilde{G}$ given by $c_n \Lambda^{\ell + \frac{3}{2}}$, with some known $c_n$, so that their contribution to the time-domain Green's function $G(t,t', r, r')$, which we denote as $G_\text{dS}(t,t',r, r')$, is given by the following residue sum, 
\be
G_\text{dS}(t,t',r, r') = -2\pi i\sum^\infty_{n=0} c_n \Lambda^{\ell + \frac{3}{2}} \frac{e^{-i \omega_n^{(\text{dS})} (t-t')}}{2\pi}.
\ee
This sum is convergent provided $t-t'+r_*+r_*'> 0$ and can be evaluated in closed form. This condition corresponds to a lightray along the lines of the region separations in \cite{Arnaudo:2025uos}. The outcome is a rational function of $e^{-\sqrt{\frac{\Lambda}{3}} (t-t'+r_*+r'_*)}$, which when expanded to leading order in $\Lambda$ gives,
\be\label{priceslawintro}
G_\text{dS}(t,t',r,r') = C R_h^{2\ell+3}(t-t'+r_*+r_*')^{-2\ell-3},
\ee
where $C$ is a dimensionless constant independent of $\Lambda$ given by \eqref{Cdef}, thus reproducing Price's law at $t\to \infty$. Note that we didn't require a large $t$ expansion to derive \eqref{priceslawintro} (just $t-t'+r_*+r_*'> 0$), so our result \eqref{priceslawintro} is exact in $t$, rather than asymptotic.

This calculation is equivalent to the integral around the branch cut of $\widetilde{G}(\omega,r,r')$ at $\Lambda = 0$. In the limit $\Lambda \to 0^+$ we show analytically that the poles at \eqref{dSmodes} coalesce to form a logarithmic cut (see the right panel of FIG. \ref{fig:analytic_structure}). This follows from a particular arrangement of gamma functions that appear in Heun connection formulae, detailed in and around \eqref{confluencecut}. One then obtains the same result \eqref{priceslawintro} from the corresponding cut discontinuity integral,
\be\label{discintro}
G_\text{dS}(t,t',r,r') = -\int_{-i \infty}^0 \text{Disc}_{\omega}\widetilde{G}(\omega,r,r') e^{-i \omega (t-t')} \frac{d\omega}{2\pi},
\ee
where
\bea
\text{Disc}_{\omega}\widetilde{G}(\omega,r,r') &\equiv& \lim_{\epsilon\to 0}\left(\widetilde{G}(\omega + \epsilon,r,r') - \widetilde{G}(\omega - \epsilon,r,r')\right)\nonumber\\
&=&  \frac{2\pi i C R_h^{2\ell+3}(i\omega)^{2\ell + 2}e^{-i \omega(r_*+r_*')}}{\Gamma(3+2\ell)} + \ldots.
\eea

The class of linear perturbations around Schwarzschild-de Sitter that we consider can be described by a single ODE having four regular singularities, which is known as the Heun equation \cite{heun1888theorie, ronveaux1995heun}.
The analytic technique we use to study the differential equation and perform the near-horizon residue computations comes from the context of Seiberg-Witten theory and supersymmetric gauge theory \cite{seiberg1994, seiberg1994a}, and Liouville conformal field theory (CFT) (see \cite{Teschner:2001rv,Nakayama:2004vk} for detailed reviews), where the Heun equation arises as the equation satisfied by a correlation function with four primary insertions plus a degenerate one \cite{Belavin:1984vu}. The relevant dual gauge theory (under AGT duality \cite{Alday:2009aq}) to Liouville CFT on a four-punctured sphere is the $\mathcal{N}=2$ $SU(2)$ supersymmetric gauge theory with $N_f=4$ fundamental hypermultiplet.
The connection between black hole perturbation theory and Seiberg-Witten curves was established in \cite{Aminov:2020yma} and was then developed in many subsequent works (see for example \cite{Bianchi:2021xpr, Bonelli:2021uvf, Bianchi:2021mft, daCunha:2022ewy, Fioravanti:2021dce, Dodelson:2022yvn, Jia:2024zes, Arnaudo:2025kof}).
In particular, the solutions to the Heun equation, as well as the connection coefficients relating the local solutions around different singular points, can be written explicitly in terms of conformal block expansions of the correlation function in the Liouville CFT language \cite{Litvinov:2013sxa, Piatek:2017fyn}, or in terms of Nekrasov partition functions \cite{Nekrasov:2002qd, Nekrasov:2003rj}.\footnote{We will use the latter language, and specifically work in the Nekrasov-Shatashvili (NS) phase of the Omega-background \cite{Nekrasov:2009rc}. We postpone to the supplemental material the conventions used for the NS functions.}
We follow the conventions set up in the recent works \cite{Bonelli:2022ten, Lisovyy2022}.

It is by using these techniques that we are able to analytically describe the tail in the near-horizon region, since in the $\Lambda \to 0^+$ limit, both the local Heun functions and the confluent Heun ones (resulting from the limiting procedure) have a finite radius of convergence there. In the far-region, instead, the strict $\Lambda\to 0^+$ limit would still need the treatment of the expansion of local solutions around the irregular singularity. Thus, in this region, we eschew the analytical treatment, and perform a brute force residue partial sum numerically at small but finite $\Lambda$, and in doing so we also obtain the Price's law tail.

\begin{figure}[t]
    \centering
    \includegraphics[width=\columnwidth]{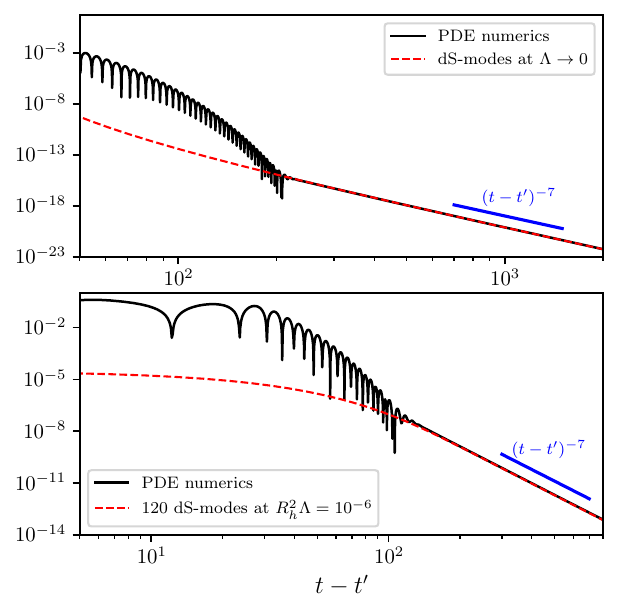}
    \caption{Price's power-law tail derived from de Sitter QNM residue sums (red, dashed) compared to a time-domain numerical solution of the Schwarzschild Green's function PDE (\eqref{GreensPDE} at $\Lambda = 0$) using Gaussian initial data of width $\sigma$ to approximate the delta function (black, solid). A power law of $(t-t')^{-2\ell - 3}$ with arbitrary normalisation is indicated in blue as reference. We use parameters $s= \ell = 2, \sigma = 1, R_h = 1$.
    \textbf{Upper panel:} Behaviour near the black hole horizon $r_* = -12.2006$, $r_*' = -10.1005$. Here, the dS mode infinite residue sum is computed analytically in the limit $\Lambda \to 0$, given by \eqref{priceslawintro}, and is shown analytically to be equal to the corresponding cut discontinuity integral. \textbf{Lower panel:} Behaviour far from the black hole, $r_*' = 12.0506$, $r_* = 16.0508$. Here, a partial dS mode residue sum is computed numerically from solutions to the Heun differential equation, at small $R_h^2\Lambda = 10^{-6}$.}
    \label{fig:numerics}
\end{figure}

Our results from both calculations (analytic near the horizon, and numerical elsewhere) agree precisely with the tail obtained from a direct numerical solution to the Schwarzschild PDE (\eqref{GreensPDE} at $\Lambda = 0$), as illustrated in FIG. \ref{fig:numerics}. 

Previously, it was pointed out that an artificial discretisation of the branch cut can also reproduce the power-law tail in Schwarzschild \cite{Besson:2024adi}.
Partial results from this study were first reported in \cite{Arnaudo:2025uos}.

\section{Schwarzschild-de Sitter and its Green's function}

We study linear perturbations around the four-dimensional Schwarzschild-de Sitter black hole 
\begin{equation}
ds^2=-f(r)\,dt^2+\frac{dr^2}{f(r)}+r^2\,d\Omega_2^2,
\end{equation}
where 
\begin{equation}\label{fr}
f(r)=1-\frac{2M}{r}-\frac{\Lambda}{3}\,r^2,
\end{equation}
with $M$ being the mass of the black hole, and $\Lambda>0$ the cosmological constant. Assuming $0<9\,\Lambda\,M^2<1$, the equation $r\,f(r)=0$ has three real roots. We denote with $R_h$ the event horizon, with $R_+>R_h$ the cosmological horizon, and with $R_-<0$ the negative root.

We consider a class of linear perturbations encoded in the following Regge-Wheeler-like ODE\footnote{The notations used here are in accordance with those in \cite{Arnaudo:2025uos}.}
\begin{equation}\label{reggewheeler}
\phi''(r) + \frac{f'(r)}{f(r)}\, \phi'(r) +\frac{\omega^2-V(r)}{f(r)^2}\,\phi(r)=0,
\end{equation}
with
\begin{equation}\label{potentialRW}
V(r)= f(r)\left[\frac{\ell(\ell+1)}{r^2}+(1-s^2)\left(\frac{3R_h-\Lambda R_h^3}{3 r^3}\right)\right].
\end{equation}
The ODE \eqref{reggewheeler} describes a conformally-coupled scalar perturbation for $s=0$\footnote{The differential equation for $s=0$ reduces to the massless scalar field perturbation in the flat limit.}, an electromagnetic perturbation for $s=1$, and vector-type gravitational perturbation for $s=2$.

By defining the variable
\begin{equation}
z=\frac{R_h}{r}\frac{r-R_-}{R_h-R_-},
\end{equation}
and redefining the wave function as 
\begin{equation}\label{wfredef}
\phi(z)=p(z)\psi(z),
\end{equation}
where $p(z)$ is given in \eqref{pz}, the ODE satisfied by $\psi(z)$ is a Heun equation \cite{heun1888theorie, ronveaux1995heun} of the form
\begin{equation}\label{heunnormalform}
\begin{aligned}
&\frac{\mathrm{d}^2\,\psi(z)}{\mathrm{d}\,z^2} + \Biggl[\frac{\frac{1}{4}-a_0^2}{z^2}+\frac{\frac{1}{4}-a_1^2}{(z-1)^2} + \frac{\frac{1}{4}-a_x^2}{(z-x)^2}\\
&\ - \frac{\frac{1}{2}-a_1^2 -a_x^2 -a_0^2 +a_\infty^2 + u}{z(z-1)}+\frac{u}{z(z-x)} \Biggr]\psi(z)=0,
\end{aligned}
\end{equation}
with dictionary given in \eqref{dictioheun}.
The black hole horizon is located at $z=1$, and the cosmological horizon is at $z=x$, which approaches $z=0$ in the $\Lambda\to 0^+$ limit: 
\begin{equation}
x=2R_h \sqrt{\frac{\Lambda}{3}} +\mathcal{O}\left(\Lambda\right).
\end{equation}
The local solutions selected by the ingoing boundary condition at the horizon and the ougoing boundary condition at the cosmological horizon are normalised as
\begin{equation}
\begin{aligned}
\psi_{\text{in}}(z)&=(1-z)^{\frac{1}{2}-a_1}\left[1+\mathcal{O}\left(1-z\right)\right],\quad z\to 1,\\
\psi_{\text{up}}(z)&=(z-x)^{\frac{1}{2}-a_x}\left[1+\mathcal{O}\left(1-z\right)\right],\quad z\to x,
\end{aligned}
\end{equation}
respectively, and can be found explicitly in \eqref{insol} and \eqref{upsol}.

The retarded Green's function in position space $G(t,t',r,r')$, at fixed values of the angular quantum numbers, is the solution to the PDE
\begin{equation}\label{GreensPDE}
\left(-\partial_t^2 + \partial_{r_*}^2 - V(r)\right)G(t,t',r,r') = -\delta(t-t')\delta(r_*-r_*'),
\end{equation}
which vanishes for $t<t'$, and where $r_*$ denotes the tortoise coordinate. In the Schwarzschild-de Sitter geometry, we use the tortoise coordinate in \eqref{tortoise}. 
The Green's function $G(t,t',r,r')$ can be constructed by introducing the Green's function $\widetilde{G}(\omega,r,r')$ in frequency space as 
\begin{equation}\label{Greensint}
G(t,t',r,r') = \int_{-\infty + i\epsilon}^{\infty + i\epsilon} \frac{d\omega}{2\pi} \widetilde{G}(\omega,r,r') e^{-i \omega (t-t')},
\end{equation}
with $\widetilde{G}(\omega,r,r')$ defined starting from the homogeneous solutions of \eqref{reggewheeler} satisfying the previously described boundary conditions as 
\begin{equation}\label{tildeGfull}
\widetilde{G}(\omega, r, r') = \frac{1}{\mathcal{W}} \times
\begin{cases}
\phi_\text{in}(r) \, \phi_\text{up}(r'), & r < r' \\
\phi_\text{in}(r') \, \phi_\text{up}(r), & r > r',
\end{cases}
\end{equation}
where 
$\mathcal{W}$ is the Wronskian $\mathcal{W} = \phi_\text{up} \partial_{r_\ast}\phi_\text{in} - \phi_\text{in} \partial_{r_\ast}\phi_\text{up}$.

By closing the integration contour of \eqref{Greensint} in the lower half-plane, the Green's function can be written as a sum of poles, the QNMs located at zeros of $\mathcal{W}$, plus the large-frequency arc contribution.
We are interested in the late-time behavior of the Green's function, where the QNM sum is convergent and the arc contribution vanishes \cite{Arnaudo:2025uos}.

\section{Analytic results near the horizon}\label{sec:inner}

Passing from a small positive $\Lambda$ to the $\Lambda=0$ case, the singularity structure of the Green's function $\widetilde{G}(\omega,r,r')$ changes. More precisely, 
analogously to the mechanism analysed in \cite{Arnaudo:2024sen} for the extremal limit of asymptotically AdS$_5$ Reissner-N\"ordstrom planar black hole, 
there is a set of QNMs that coalesce in a branch cut in the $\Lambda\to 0^+$ limit.
To investigate this structure, we analytically continue one of the local solutions $\psi_{\text{in}},\psi_{\text{up}}$ to the region where the other one converges. Since we want also to consider the $\Lambda\to 0^+$ limit, our choice is to connect to the near-horizon region, where the local solutions have a finite radius of convergence also at $\Lambda=0$. In other words, we write
\begin{equation}
\psi_\text{up}(z) = C_{\text{up},\text{in}}\,\psi_\text{in}(z)+C_{\text{up},\text{out}}\,\psi_\text{out}(z).
\end{equation}
The explicit expressions for $\psi_{\text{out}}(z)$, $C_{\text{up},\text{in}}$, and $C_{\text{up},\text{out}}$ can be found in \eqref{psiout},\eqref{Cupin},\eqref{Cupout}. 

As a result of the connection procedure, the Green's function $\widetilde{G}(\omega, r, r')$ is split into two terms, only one of which has poles at the QNMs:
\begin{equation}\label{Gplus}
\begin{aligned}
\widetilde{G}_+(\omega,z,z')=\frac{-3 R_h R_-}{\Lambda\,(R_h-R_-)}p(z)p(z')\,\frac{C_{\text{up},\text{in}}}{C_{\text{up},\text{out}}}\frac{\psi_{\mathrm{in}}(z)\,\psi_{\mathrm{in}}(z')}{2a_1}.
\end{aligned}    
\end{equation}
The full expression for the ratio of the connection coefficients can be organised as in \eqref{ratioconncoeff}.

The set of poles of \eqref{Gplus} includes a branch of longer-living de Sitter modes, which can be found in the small $\Lambda$ regime as poles of the $\Gamma$ function $\Gamma\left(\frac{1}{2}+a-a_x+a_0\right)$:
\begin{equation}
\frac{1}{2}+a-a_{x}+a_0=-n,\quad n\in\mathbb{Z}_{\ge 0},\label{quantcond}
\end{equation}
from which we obtain the mode frequencies \eqref{dSmodes}.
Here, $a$ parametrises the composite monodromy around the points $z=0$ and $z=x$. 
The parameter $a$ is defined as an instanton expansion in powers of $x$ as explained in the Supplemental Material, and the leading order in the instanton expansion can be found in \eqref{a0inst}.
When the frequency is of order $\sqrt{\Lambda}$ in the small $\Lambda$ expansion, as for the modes \eqref{dSmodes}, the small $\Lambda$ expansion of $a$ reads 
\begin{equation}\label{aexp}
a=\ell+\frac{1}{2}+a_{\Lambda}^{(1)}\,\Lambda+\mathcal{O}\left(\Lambda^{3/2}\right),
\end{equation}
where $a_{\Lambda}^{(1)}$ is given in \eqref{a1}.

In the $\Lambda\to 0^+$ limit, $a_0-a_x\to\infty$ and $x\to 0$ with $x\left(a_0-a_x\right)$ finite.\footnote{In the gauge theory language, this is the decoupling limit of a fundamental multiplet with mass $a_0-a_x$, that leads to a $N_f=3$ theory starting from the $N_f=4$ one.} As a consequence, the set of modes in \eqref{dSmodes} coalesces into a branch cut, which can be seen explicitly inside \eqref{greensparts2} from the limiting procedure
\begin{equation}\label{confluencecut}
\begin{aligned}
&x^{2a}\frac{\Gamma\left(\frac{1}{2}+a-a_x+a_0\right)}{\Gamma\left(\frac{1}{2}-a-a_x+a_0\right)}=\\
&x^{2a}\left(a_0-a_x\right)^{2a}\left[1+\mathcal{O}\left(\frac{1}{a_0-a_x}\right)\right]\to \left(-2 i \omega R_h\right)^{2a},
\end{aligned}
\end{equation}
where we used
\begin{equation}
\frac{\Gamma(z+\alpha)}{\Gamma(z+\beta)}= z^{\alpha-\beta}\left[1+\mathcal{O}\left(\frac{1}{z}\right)\right],
\end{equation}
for the first equality, and then we took the $\Lambda\to 0^+$ limit. \eqref{confluencecut} has a logarithmic branch point at the origin, as can be seen by expanding at small $\omega$,
\bea
\left(-2 i \omega R_h\right)^{2a} &=& \frac{3\,(-2 i R_h \omega )^{2 \ell+3}}{2 (2 \ell+1)}\log\left(-2 i\omega R_h\right)\nonumber\\
&&+ (\text{terms regular at $\omega = 0$}) \\
&&+ (\text{subleading terms})\nonumber
\eea
with the cut along the negative imaginary axis. The full expression is given in \eqref{logcut}.

From the black hole perturbation point of view, this branch cut emerging in the $\Lambda\to 0^+$ limit is known to be responsible for the late-time power-law decay of the retarded Green's function $G(t,t',r,r')$. Here, we prove that the same power-law can be recovered from a sum over the residues of the poles \eqref{dSmodes} in \eqref{Gplus}.

We begin by expanding the ratio of the connection coefficients \eqref{ratioconncoeff} in the small $x$ regime, obtaining the expression \eqref{ratioexp}.\footnote{Here, we are also working at leading order in the instanton expansion. We postpone to the supplemental material the discussion of what this expansion looks like and which conventions we are using.}
We note that the frequencies \eqref{dSmodes} are not poles of the leading order term in the first line of \eqref{ratioexp}, but of the subleading one, scaling with $x^{2a}$.

To simplify the analytic computation, we work in the asymptotic $r_*,r'_*\to -\infty$ regime, and replace the wave functions 
\begin{equation}\label{replacement}
\tilde{\psi}(z)\equiv \rho_h\,p(z)\,\psi_{\text{in}}(z),
\end{equation}
where $\rho_h$ can be found in \eqref{rhoh}, with $e^{-i\,\omega\,r_*(z)}$.
By using the quantisation condition \eqref{quantcond}, the contribution of the set of de Sitter modes \eqref{dSmodes} to the Green's function is given by the sum of residues
\begin{widetext}
\begin{equation}\label{residuesum}
\begin{aligned}
G_\text{dS}(t,t',r,r') =2\pi i\,\sum_{n\ge 0}\frac{3 R_h R_-}{\Lambda\,(R_h-R_-)}\frac{e^{-i\omega(t-t'+r_*+r'_*)}}{2\pi}\frac{x^{2a}}{2a_1\,\rho_h^2}\,\mathfrak{G}\,\frac{(-1)^{n}}{n!}\left(\frac{d(a+a_0-a_x)}{d \omega }\right)^{-1}\bigg|_{\omega=\omega_n^{\text{dS}}},
\end{aligned}
\end{equation}
\end{widetext}
where the expression of $\mathfrak{G}$ can be found in \eqref{frakG}. 

By using the small $\Lambda$ expansions of quantities appearing in \eqref{residuesum}, i.e. \eqref{aexp} \eqref{expansioncoeff}, we arrive at
\bea\label{simplifiedsum}
&&G_\text{dS}(t,t',r,r') = C R_h^{2\ell+3}\times\\
&&\quad\frac{\Lambda^{\ell+\frac{3}{2}}}{2\,3^{\ell+\frac{3}{2}}\,(\ell+1)}\sum_{j=\ell+1}^{\infty}e^{- \sqrt{\frac{\Lambda}{3}} (t-t'+r_*+r'_*)\,j}j\, \binom{j+ \ell}{2\ell+1},\nonumber
\eea
where $C$ is defined in \eqref{Cdef}.
The series appearing in \eqref{simplifiedsum} is a power series with polynomial coefficients, which can be computed explicitly to give the following rational function,
\be
\begin{aligned}
\frac{(\ell+1)\left(e^{\frac{\sqrt{\Lambda } (t-t'+r_*+r'_*)}{\sqrt{3}}}+1\right) e^{-\frac{\sqrt{\Lambda } (\ell-1) (t-t'+r_*+r'_*)}{\sqrt{3}}} }{\left(e^{\frac{\sqrt{\Lambda } (t-t'+r_*+r'_*)}{\sqrt{3}}}-1\right)^3\left(1-e^{-\frac{\sqrt{\Lambda } (t-t'+r_*+r'_*)}{\sqrt{3}}}\right)^{2 \ell}}.
\end{aligned}
\ee
Taking the $\Lambda\to 0^+$ limit of the resulting $G_{\text{dS}}$, we find Price's law, \eqref{priceslawintro}, including the coefficient. As shown in FIG. \ref{fig:numerics} this result matches precisely the tail obtained from a numerical solution to the PDE \eqref{GreensPDE}, again, including the coefficient.

We can alternatively get the same result, \eqref{priceslawintro}, by performing a discontinuity integral of the branch cut which emerges at $\Lambda \to 0$ via the limiting procedure in \eqref{confluencecut}.\footnote{We show this result by again expanding all the expressions at zero instantons.} Indeed, using the substitutions \eqref{confluencecut} and \eqref{replacement}, we can write $\widetilde{G}_+(\omega,z,z')$ as
\begin{equation}
\begin{aligned}
\widetilde{G}_+(\omega,z,z')=\frac{-3 R_h R_-}{\Lambda\,(R_h-R_-)}\,\frac{\tilde{C}_{\text{up},\text{in}}}{\tilde{C}_{\text{up},\text{out}}}\frac{e^{-i\omega(r_*(z)+r'_*(z'))}}{2a_1\,\rho_h^2},
\end{aligned}
\end{equation}
with $\tilde{C}_{\text{up},\text{in}}/\tilde{C}_{\text{up},\text{out}}$ given in \eqref{ratioconncoeffcut}. 
The branch cut structure can be seen explicitly from the logarithm in \eqref{logcut}. By using the small $\Lambda$-expansions and the small $\omega$-expansions in \eqref{coeffcutomegaexp1} and \eqref{coeffcutomegaexp2}, the computation of $G_\text{dS}(t,t',r,r')$ reduces to the discontinuity integral \eqref{discintro}, which again gives Price's law \eqref{priceslawintro}.

\section{Numerical results for generic $r_*, r_*'$}
In the previous section, we derived the tail of the Schwarzschild Green's function when $r_*,r'_*$ are close to the black hole horizon, as a QNM sum, using analytic techniques. In this section, we address the same computation for generic $r_*, r_*'$, in particular, not near the horizon. In this case the analytical treatment in the $\Lambda\to 0^+$ limit is complicated by the irregular singularity. Instead, we utilise a numerical approach at small, but finite $\Lambda$, and sum the first $N$ de Sitter mode residues.

The full expression \eqref{tildeGfull}, including $\mathcal{W}$, can be expressed in terms of Heun functions, $\psi_\text{in}(z)$ and $\psi_\text{up}(z)$, as given in \eqref{insol} and \eqref{upsol} and their derivatives. At finite $\Lambda$ these can be numerically evaluated efficiently.\footnote{We use \texttt{HeunG} in Mathematica.} 

First, we evaluate the de Sitter mode frequencies, $\omega_n^\text{dS}$, which are roots of $\mathcal{W}$. At small $\Lambda$, the $\omega_n^\text{dS}$ are approximated by \eqref{dSmodes}, which serve as an initial guess for root finding using Newton's method. 

Once $\omega_n^\text{dS}$ are obtained for $n = 0, \ldots, N-1$, we select our chosen values for $r, r'$ and numerically evaluate
\be
\text{res}_n \equiv \text{res}\left(\widetilde{G}(\omega, r,r') \frac{e^{-i \omega (t-t')}}{2\pi}, \omega_n^\text{dS}\right),
\ee
for each $n$.
Practically, this can be achieved by first computing the gradient of $\mathcal{W}$ at the root through a numerical derivative,
\be
\mathcal{S}^{\epsilon}_n =\frac{\mathcal{W}(\omega_n^\text{dS} + \epsilon) - \mathcal{W}(\omega_n^\text{dS})}{\epsilon},
\ee
for some small $\epsilon$. The residue then follows by evaluating the remaining regular pieces of $\widetilde{G}$ and combining with the gradient of $\mathcal{W}$, i.e.
\be
\text{res}^{\epsilon}_n = \frac{\left(\widetilde{G}(\omega, r,r')\,\mathcal{W}\, \frac{e^{-i \omega (t-t')}}{2\pi}\right)\big|_{\omega=\omega_n^\text{(dS)}}}{\mathcal{S}^{\epsilon}_n}.
\ee
Finally we compute the partial sum
$G_\text{dS}^{(N)} = \sum_{n=0}^{N-1}\text{res}^{\epsilon}_n$,
which is now a function of $t-t'$ with numerical coefficients.

Let us denote the longest-lived Schwarzschild QNM as $\omega^\text{S}_0$. In the case of $s=2,\ell=2$ we have $\omega^\text{S}_0 \simeq (\pm 0.747343 - 0.177925i)R_h^{-1}$ (see e.g. \cite{Leaver:1985ax}). For the de Sitter modes to dominate at late times, one therefore requires $\Lambda R_h^2 \lesssim 0.0105524$.
Provided $\Lambda R_h^2$ obeys this inequality, $G(t, t', r, r')$ will decay exponentially at asymptotically late times by the longest lived de Sitter mode rate. However, at small enough $\Lambda R_h^2$ and with large enough $N$, $G_\text{dS}^{(N)}$ will start to resemble the power-law tail. Longer tails require smaller $\Lambda R_h^2$ and larger $N$, and in the limit $\Lambda \to 0$ the tail becomes the asymptotic behaviour. 

In the lower panel of FIG. \ref{fig:numerics} we show $G_\text{dS}^{(N)}$ with $N=120$ and $\Lambda R_h^2 = 10^{-6}$, demonstrating the emergence of the tail over the timescales shown. We ensure the roots $\omega_n^\text{dS}$ are converged to $10^{-100}$, and we use $\epsilon = 10^{-60}$.

\paragraph{Acknowledgements.}
\begin{acknowledgments}
It is a pleasure to thank Javier Carballo for discussions.
PA is supported by the Royal Society grant URF\textbackslash R\textbackslash 231002, `Dynamics of holographic field theories'.
BW is supported by a Royal Society University Research Fellowship and in part by the Science and Technology Facilities Council (Consolidated Grant `New Frontiers in Particle Physics, Cosmology and Gravity'). 
\end{acknowledgments}

\bibliography{biblio} 

\begin{thebibliography}{41}%
\makeatletter
\providecommand \@ifxundefined [1]{%
 \@ifx{#1\undefined}
}%
\providecommand \@ifnum [1]{%
 \ifnum #1\expandafter \@firstoftwo
 \else \expandafter \@secondoftwo
 \fi
}%
\providecommand \@ifx [1]{%
 \ifx #1\expandafter \@firstoftwo
 \else \expandafter \@secondoftwo
 \fi
}%
\providecommand \natexlab [1]{#1}%
\providecommand \enquote  [1]{``#1''}%
\providecommand \bibnamefont  [1]{#1}%
\providecommand \bibfnamefont [1]{#1}%
\providecommand \citenamefont [1]{#1}%
\providecommand \href@noop [0]{\@secondoftwo}%
\providecommand \href [0]{\begingroup \@sanitize@url \@href}%
\providecommand \@href[1]{\@@startlink{#1}\@@href}%
\providecommand \@@href[1]{\endgroup#1\@@endlink}%
\providecommand \@sanitize@url [0]{\catcode `\\12\catcode `\$12\catcode `\&12\catcode `\#12\catcode `\^12\catcode `\_12\catcode `\%12\relax}%
\providecommand \@@startlink[1]{}%
\providecommand \@@endlink[0]{}%
\providecommand \url  [0]{\begingroup\@sanitize@url \@url }%
\providecommand \@url [1]{\endgroup\@href {#1}{\urlprefix }}%
\providecommand \urlprefix  [0]{URL }%
\providecommand \Eprint [0]{\href }%
\providecommand \doibase [0]{http://dx.doi.org/}%
\providecommand \selectlanguage [0]{\@gobble}%
\providecommand \bibinfo  [0]{\@secondoftwo}%
\providecommand \bibfield  [0]{\@secondoftwo}%
\providecommand \translation [1]{[#1]}%
\providecommand \BibitemOpen [0]{}%
\providecommand \bibitemStop [0]{}%
\providecommand \bibitemNoStop [0]{.\EOS\space}%
\providecommand \EOS [0]{\spacefactor3000\relax}%
\providecommand \BibitemShut  [1]{\csname bibitem#1\endcsname}%
\let\auto@bib@innerbib\@empty
\bibitem [{\citenamefont {Price}(1972{\natexlab{a}})}]{Price1}%
  \BibitemOpen
  \bibfield  {author} {\bibinfo {author} {\bibfnamefont {R.~H.}\ \bibnamefont {Price}},\ }\href {\doibase 10.1103/PhysRevD.5.2419} {\bibfield  {journal} {\bibinfo  {journal} {Phys. Rev. D}\ }\textbf {\bibinfo {volume} {5}},\ \bibinfo {pages} {2419} (\bibinfo {year} {1972}{\natexlab{a}})}\BibitemShut {NoStop}%
\bibitem [{\citenamefont {Price}(1972{\natexlab{b}})}]{Price2}%
  \BibitemOpen
  \bibfield  {author} {\bibinfo {author} {\bibfnamefont {R.~H.}\ \bibnamefont {Price}},\ }\href {\doibase 10.1103/PhysRevD.5.2439} {\bibfield  {journal} {\bibinfo  {journal} {Phys. Rev. D}\ }\textbf {\bibinfo {volume} {5}},\ \bibinfo {pages} {2439} (\bibinfo {year} {1972}{\natexlab{b}})}\BibitemShut {NoStop}%
\bibitem [{\citenamefont {Donninger}\ \emph {et~al.}(2011)\citenamefont {Donninger}, \citenamefont {Schlag},\ and\ \citenamefont {Soffer}}]{DONNINGER2011484}%
  \BibitemOpen
  \bibfield  {author} {\bibinfo {author} {\bibfnamefont {R.}~\bibnamefont {Donninger}}, \bibinfo {author} {\bibfnamefont {W.}~\bibnamefont {Schlag}}, \ and\ \bibinfo {author} {\bibfnamefont {A.}~\bibnamefont {Soffer}},\ }\href {\doibase https://doi.org/10.1016/j.aim.2010.06.026} {\bibfield  {journal} {\bibinfo  {journal} {Advances in Mathematics}\ }\textbf {\bibinfo {volume} {226}},\ \bibinfo {pages} {484} (\bibinfo {year} {2011})}\BibitemShut {NoStop}%
\bibitem [{\citenamefont {Price}\ and\ \citenamefont {Burko}(2004)}]{Price:2004mm}%
  \BibitemOpen
  \bibfield  {author} {\bibinfo {author} {\bibfnamefont {R.~H.}\ \bibnamefont {Price}}\ and\ \bibinfo {author} {\bibfnamefont {L.~M.}\ \bibnamefont {Burko}},\ }\href {\doibase 10.1103/PhysRevD.70.084039} {\bibfield  {journal} {\bibinfo  {journal} {Phys. Rev. D}\ }\textbf {\bibinfo {volume} {70}},\ \bibinfo {pages} {084039} (\bibinfo {year} {2004})},\ \Eprint {http://arxiv.org/abs/gr-qc/0408077} {arXiv:gr-qc/0408077} \BibitemShut {NoStop}%
\bibitem [{\citenamefont {Leaver}(1986)}]{leaver1986}%
  \BibitemOpen
  \bibfield  {author} {\bibinfo {author} {\bibfnamefont {E.~W.}\ \bibnamefont {Leaver}},\ }\href {\doibase 10.1103/PhysRevD.34.384} {\bibfield  {journal} {\bibinfo  {journal} {Phys. Rev. D}\ }\textbf {\bibinfo {volume} {34}},\ \bibinfo {pages} {384} (\bibinfo {year} {1986})}\BibitemShut {NoStop}%
\bibitem [{\citenamefont {Ching}\ \emph {et~al.}(1995)\citenamefont {Ching}, \citenamefont {Leung}, \citenamefont {Suen},\ and\ \citenamefont {Young}}]{ching-PhysRevD.52.2118}%
  \BibitemOpen
  \bibfield  {author} {\bibinfo {author} {\bibfnamefont {E.~S.~C.}\ \bibnamefont {Ching}}, \bibinfo {author} {\bibfnamefont {P.~T.}\ \bibnamefont {Leung}}, \bibinfo {author} {\bibfnamefont {W.~M.}\ \bibnamefont {Suen}}, \ and\ \bibinfo {author} {\bibfnamefont {K.}~\bibnamefont {Young}},\ }\href {\doibase 10.1103/PhysRevD.52.2118} {\bibfield  {journal} {\bibinfo  {journal} {Phys. Rev. D}\ }\textbf {\bibinfo {volume} {52}},\ \bibinfo {pages} {2118} (\bibinfo {year} {1995})}\BibitemShut {NoStop}%
\bibitem [{\citenamefont {Andersson}(1997)}]{Andersson:1996cm}%
  \BibitemOpen
  \bibfield  {author} {\bibinfo {author} {\bibfnamefont {N.}~\bibnamefont {Andersson}},\ }\href {\doibase 10.1103/PhysRevD.55.468} {\bibfield  {journal} {\bibinfo  {journal} {Phys. Rev. D}\ }\textbf {\bibinfo {volume} {55}},\ \bibinfo {pages} {468} (\bibinfo {year} {1997})},\ \Eprint {http://arxiv.org/abs/gr-qc/9607064} {arXiv:gr-qc/9607064} \BibitemShut {NoStop}%
\bibitem [{\citenamefont {Maassen van~den Brink}(2004)}]{MaassenvandenBrink:2003as}%
  \BibitemOpen
  \bibfield  {author} {\bibinfo {author} {\bibfnamefont {A.}~\bibnamefont {Maassen van~den Brink}},\ }\href {\doibase 10.1063/1.1626805} {\bibfield  {journal} {\bibinfo  {journal} {J. Math. Phys.}\ }\textbf {\bibinfo {volume} {45}},\ \bibinfo {pages} {327} (\bibinfo {year} {2004})},\ \Eprint {http://arxiv.org/abs/gr-qc/0303095} {arXiv:gr-qc/0303095} \BibitemShut {NoStop}%
\bibitem [{\citenamefont {Casals}\ and\ \citenamefont {Ottewill}(2012{\natexlab{a}})}]{Casals:2011aa}%
  \BibitemOpen
  \bibfield  {author} {\bibinfo {author} {\bibfnamefont {M.}~\bibnamefont {Casals}}\ and\ \bibinfo {author} {\bibfnamefont {A.}~\bibnamefont {Ottewill}},\ }\href {\doibase 10.1103/PhysRevD.86.024021} {\bibfield  {journal} {\bibinfo  {journal} {Phys. Rev. D}\ }\textbf {\bibinfo {volume} {86}},\ \bibinfo {pages} {024021} (\bibinfo {year} {2012}{\natexlab{a}})},\ \Eprint {http://arxiv.org/abs/1112.2695} {arXiv:1112.2695 [gr-qc]} \BibitemShut {NoStop}%
\bibitem [{\citenamefont {Casals}\ and\ \citenamefont {Ottewill}(2012{\natexlab{b}})}]{casals1}%
  \BibitemOpen
  \bibfield  {author} {\bibinfo {author} {\bibfnamefont {M.}~\bibnamefont {Casals}}\ and\ \bibinfo {author} {\bibfnamefont {A.}~\bibnamefont {Ottewill}},\ }\href {\doibase 10.1103/PhysRevLett.109.111101} {\bibfield  {journal} {\bibinfo  {journal} {Phys. Rev. Lett.}\ }\textbf {\bibinfo {volume} {109}},\ \bibinfo {pages} {111101} (\bibinfo {year} {2012}{\natexlab{b}})}\BibitemShut {NoStop}%
\bibitem [{\citenamefont {Casals}\ and\ \citenamefont {Ottewill}(2013)}]{Casals:2012ng}%
  \BibitemOpen
  \bibfield  {author} {\bibinfo {author} {\bibfnamefont {M.}~\bibnamefont {Casals}}\ and\ \bibinfo {author} {\bibfnamefont {A.~C.}\ \bibnamefont {Ottewill}},\ }\href {\doibase 10.1103/PhysRevD.87.064010} {\bibfield  {journal} {\bibinfo  {journal} {Phys. Rev. D}\ }\textbf {\bibinfo {volume} {87}},\ \bibinfo {pages} {064010} (\bibinfo {year} {2013})},\ \Eprint {http://arxiv.org/abs/1210.0519} {arXiv:1210.0519 [gr-qc]} \BibitemShut {NoStop}%
\bibitem [{\citenamefont {Casals}\ and\ \citenamefont {Ottewill}(2015)}]{Casals:2015nja}%
  \BibitemOpen
  \bibfield  {author} {\bibinfo {author} {\bibfnamefont {M.}~\bibnamefont {Casals}}\ and\ \bibinfo {author} {\bibfnamefont {A.~C.}\ \bibnamefont {Ottewill}},\ }\href {\doibase 10.1103/PhysRevD.92.124055} {\bibfield  {journal} {\bibinfo  {journal} {Phys. Rev. D}\ }\textbf {\bibinfo {volume} {92}},\ \bibinfo {pages} {124055} (\bibinfo {year} {2015})},\ \Eprint {http://arxiv.org/abs/1509.04702} {arXiv:1509.04702 [gr-qc]} \BibitemShut {NoStop}%
\bibitem [{\citenamefont {Arnaudo}\ and\ \citenamefont {Withers}(2025)}]{Arnaudo:2024sen}%
  \BibitemOpen
  \bibfield  {author} {\bibinfo {author} {\bibfnamefont {P.}~\bibnamefont {Arnaudo}}\ and\ \bibinfo {author} {\bibfnamefont {B.}~\bibnamefont {Withers}},\ }\href {\doibase 10.1103/8n3f-2d33} {\bibfield  {journal} {\bibinfo  {journal} {Phys. Rev. D}\ }\textbf {\bibinfo {volume} {111}},\ \bibinfo {pages} {L121903} (\bibinfo {year} {2025})},\ \Eprint {http://arxiv.org/abs/2412.01923} {arXiv:2412.01923 [hep-th]} \BibitemShut {NoStop}%
\bibitem [{\citenamefont {Arnaudo}\ \emph {et~al.}(2025{\natexlab{a}})\citenamefont {Arnaudo}, \citenamefont {Carballo},\ and\ \citenamefont {Withers}}]{Arnaudo:2025uos}%
  \BibitemOpen
  \bibfield  {author} {\bibinfo {author} {\bibfnamefont {P.}~\bibnamefont {Arnaudo}}, \bibinfo {author} {\bibfnamefont {J.}~\bibnamefont {Carballo}}, \ and\ \bibinfo {author} {\bibfnamefont {B.}~\bibnamefont {Withers}},\ }\href@noop {} {\  (\bibinfo {year} {2025}{\natexlab{a}})},\ \Eprint {http://arxiv.org/abs/2510.18956} {arXiv:2510.18956 [gr-qc]} \BibitemShut {NoStop}%
\bibitem [{\citenamefont {Heun}(1888)}]{heun1888theorie}%
  \BibitemOpen
  \bibfield  {author} {\bibinfo {author} {\bibfnamefont {K.}~\bibnamefont {Heun}},\ }\href@noop {} {\bibfield  {journal} {\bibinfo  {journal} {Mathematische Annalen}\ }\textbf {\bibinfo {volume} {33}},\ \bibinfo {pages} {161} (\bibinfo {year} {1888})}\BibitemShut {NoStop}%
\bibitem [{\citenamefont {Ronveaux}\ and\ \citenamefont {Arscott}(1995)}]{ronveaux1995heun}%
  \BibitemOpen
  \bibfield  {author} {\bibinfo {author} {\bibfnamefont {A.}~\bibnamefont {Ronveaux}}\ and\ \bibinfo {author} {\bibfnamefont {F.}~\bibnamefont {Arscott}},\ }\href {https://books.google.ch/books?id=5p65FD8caCgC} {\emph {\bibinfo {title} {Heun's Differential Equations}}},\ Oxford science publications\ (\bibinfo  {publisher} {Oxford University Press},\ \bibinfo {year} {1995})\BibitemShut {NoStop}%
\bibitem [{\citenamefont {Seiberg}\ and\ \citenamefont {Witten}(1994{\natexlab{a}})}]{seiberg1994}%
  \BibitemOpen
  \bibfield  {author} {\bibinfo {author} {\bibfnamefont {N.}~\bibnamefont {Seiberg}}\ and\ \bibinfo {author} {\bibfnamefont {E.}~\bibnamefont {Witten}},\ }\href@noop {} {\bibfield  {journal} {\bibinfo  {journal} {Nucl. Phys. B}\ }\textbf {\bibinfo {volume} {431}},\ \bibinfo {pages} {484} (\bibinfo {year} {1994}{\natexlab{a}})}\BibitemShut {NoStop}%
\bibitem [{\citenamefont {Seiberg}\ and\ \citenamefont {Witten}(1994{\natexlab{b}})}]{seiberg1994a}%
  \BibitemOpen
  \bibfield  {author} {\bibinfo {author} {\bibfnamefont {N.}~\bibnamefont {Seiberg}}\ and\ \bibinfo {author} {\bibfnamefont {E.}~\bibnamefont {Witten}},\ }\href@noop {} {\bibfield  {journal} {\bibinfo  {journal} {Nucl. Phys. B}\ }\textbf {\bibinfo {volume} {426}},\ \bibinfo {pages} {19} (\bibinfo {year} {1994}{\natexlab{b}})}\BibitemShut {NoStop}%
\bibitem [{\citenamefont {Teschner}(2001)}]{Teschner:2001rv}%
  \BibitemOpen
  \bibfield  {author} {\bibinfo {author} {\bibfnamefont {J.}~\bibnamefont {Teschner}},\ }\href {\doibase 10.1088/0264-9381/18/23/201} {\bibfield  {journal} {\bibinfo  {journal} {Class. Quant. Grav.}\ }\textbf {\bibinfo {volume} {18}},\ \bibinfo {pages} {R153} (\bibinfo {year} {2001})},\ \Eprint {http://arxiv.org/abs/hep-th/0104158} {arXiv:hep-th/0104158} \BibitemShut {NoStop}%
\bibitem [{\citenamefont {Nakayama}(2004)}]{Nakayama:2004vk}%
  \BibitemOpen
  \bibfield  {author} {\bibinfo {author} {\bibfnamefont {Y.}~\bibnamefont {Nakayama}},\ }\href {\doibase 10.1142/S0217751X04019500} {\bibfield  {journal} {\bibinfo  {journal} {Int. J. Mod. Phys. A}\ }\textbf {\bibinfo {volume} {19}},\ \bibinfo {pages} {2771} (\bibinfo {year} {2004})},\ \Eprint {http://arxiv.org/abs/hep-th/0402009} {arXiv:hep-th/0402009} \BibitemShut {NoStop}%
\bibitem [{\citenamefont {Belavin}\ \emph {et~al.}(1984)\citenamefont {Belavin}, \citenamefont {Polyakov},\ and\ \citenamefont {Zamolodchikov}}]{Belavin:1984vu}%
  \BibitemOpen
  \bibfield  {author} {\bibinfo {author} {\bibfnamefont {A.~A.}\ \bibnamefont {Belavin}}, \bibinfo {author} {\bibfnamefont {A.~M.}\ \bibnamefont {Polyakov}}, \ and\ \bibinfo {author} {\bibfnamefont {A.~B.}\ \bibnamefont {Zamolodchikov}},\ }\href {\doibase 10.1016/0550-3213(84)90052-X} {\bibfield  {journal} {\bibinfo  {journal} {Nucl. Phys. B}\ }\textbf {\bibinfo {volume} {241}},\ \bibinfo {pages} {333} (\bibinfo {year} {1984})}\BibitemShut {NoStop}%
\bibitem [{\citenamefont {Alday}\ \emph {et~al.}(2010)\citenamefont {Alday}, \citenamefont {Gaiotto},\ and\ \citenamefont {Tachikawa}}]{Alday:2009aq}%
  \BibitemOpen
  \bibfield  {author} {\bibinfo {author} {\bibfnamefont {L.~F.}\ \bibnamefont {Alday}}, \bibinfo {author} {\bibfnamefont {D.}~\bibnamefont {Gaiotto}}, \ and\ \bibinfo {author} {\bibfnamefont {Y.}~\bibnamefont {Tachikawa}},\ }\href {\doibase 10.1007/s11005-010-0369-5} {\bibfield  {journal} {\bibinfo  {journal} {Lett. Math. Phys.}\ }\textbf {\bibinfo {volume} {91}},\ \bibinfo {pages} {167} (\bibinfo {year} {2010})},\ \Eprint {http://arxiv.org/abs/0906.3219} {arXiv:0906.3219 [hep-th]} \BibitemShut {NoStop}%
\bibitem [{\citenamefont {Aminov}\ \emph {et~al.}(2022)\citenamefont {Aminov}, \citenamefont {Grassi},\ and\ \citenamefont {Hatsuda}}]{Aminov:2020yma}%
  \BibitemOpen
  \bibfield  {author} {\bibinfo {author} {\bibfnamefont {G.}~\bibnamefont {Aminov}}, \bibinfo {author} {\bibfnamefont {A.}~\bibnamefont {Grassi}}, \ and\ \bibinfo {author} {\bibfnamefont {Y.}~\bibnamefont {Hatsuda}},\ }\href {\doibase 10.1007/s00023-021-01137-x} {\bibfield  {journal} {\bibinfo  {journal} {Annales Henri Poincare}\ }\textbf {\bibinfo {volume} {23}},\ \bibinfo {pages} {1951} (\bibinfo {year} {2022})},\ \Eprint {http://arxiv.org/abs/2006.06111} {arXiv:2006.06111 [hep-th]} \BibitemShut {NoStop}%
\bibitem [{\citenamefont {Bianchi}\ \emph {et~al.}(2022{\natexlab{a}})\citenamefont {Bianchi}, \citenamefont {Consoli}, \citenamefont {Grillo},\ and\ \citenamefont {Morales}}]{Bianchi:2021xpr}%
  \BibitemOpen
  \bibfield  {author} {\bibinfo {author} {\bibfnamefont {M.}~\bibnamefont {Bianchi}}, \bibinfo {author} {\bibfnamefont {D.}~\bibnamefont {Consoli}}, \bibinfo {author} {\bibfnamefont {A.}~\bibnamefont {Grillo}}, \ and\ \bibinfo {author} {\bibfnamefont {J.~F.}\ \bibnamefont {Morales}},\ }\href {\doibase 10.1016/j.physletb.2021.136837} {\bibfield  {journal} {\bibinfo  {journal} {Phys. Lett. B}\ }\textbf {\bibinfo {volume} {824}},\ \bibinfo {pages} {136837} (\bibinfo {year} {2022}{\natexlab{a}})},\ \Eprint {http://arxiv.org/abs/2105.04245} {arXiv:2105.04245 [hep-th]} \BibitemShut {NoStop}%
\bibitem [{\citenamefont {Bonelli}\ \emph {et~al.}(2022)\citenamefont {Bonelli}, \citenamefont {Iossa}, \citenamefont {Lichtig},\ and\ \citenamefont {Tanzini}}]{Bonelli:2021uvf}%
  \BibitemOpen
  \bibfield  {author} {\bibinfo {author} {\bibfnamefont {G.}~\bibnamefont {Bonelli}}, \bibinfo {author} {\bibfnamefont {C.}~\bibnamefont {Iossa}}, \bibinfo {author} {\bibfnamefont {D.~P.}\ \bibnamefont {Lichtig}}, \ and\ \bibinfo {author} {\bibfnamefont {A.}~\bibnamefont {Tanzini}},\ }\href {\doibase 10.1103/PhysRevD.105.044047} {\bibfield  {journal} {\bibinfo  {journal} {Phys. Rev. D}\ }\textbf {\bibinfo {volume} {105}},\ \bibinfo {pages} {044047} (\bibinfo {year} {2022})},\ \Eprint {http://arxiv.org/abs/2105.04483} {arXiv:2105.04483 [hep-th]} \BibitemShut {NoStop}%
\bibitem [{\citenamefont {Bianchi}\ \emph {et~al.}(2022{\natexlab{b}})\citenamefont {Bianchi}, \citenamefont {Consoli}, \citenamefont {Grillo},\ and\ \citenamefont {Morales}}]{Bianchi:2021mft}%
  \BibitemOpen
  \bibfield  {author} {\bibinfo {author} {\bibfnamefont {M.}~\bibnamefont {Bianchi}}, \bibinfo {author} {\bibfnamefont {D.}~\bibnamefont {Consoli}}, \bibinfo {author} {\bibfnamefont {A.}~\bibnamefont {Grillo}}, \ and\ \bibinfo {author} {\bibfnamefont {J.~F.}\ \bibnamefont {Morales}},\ }\href {\doibase 10.1007/JHEP01(2022)024} {\bibfield  {journal} {\bibinfo  {journal} {JHEP}\ }\textbf {\bibinfo {volume} {01}},\ \bibinfo {pages} {024} (\bibinfo {year} {2022}{\natexlab{b}})},\ \Eprint {http://arxiv.org/abs/2109.09804} {arXiv:2109.09804 [hep-th]} \BibitemShut {NoStop}%
\bibitem [{\citenamefont {da~Cunha}\ and\ \citenamefont {Cavalcante}(2024)}]{daCunha:2022ewy}%
  \BibitemOpen
  \bibfield  {author} {\bibinfo {author} {\bibfnamefont {B.~C.}\ \bibnamefont {da~Cunha}}\ and\ \bibinfo {author} {\bibfnamefont {J.~P.}\ \bibnamefont {Cavalcante}},\ }\href {\doibase 10.1007/JHEP08(2024)110} {\bibfield  {journal} {\bibinfo  {journal} {JHEP}\ }\textbf {\bibinfo {volume} {08}},\ \bibinfo {pages} {110} (\bibinfo {year} {2024})},\ \Eprint {http://arxiv.org/abs/2211.03551} {arXiv:2211.03551 [hep-th]} \BibitemShut {NoStop}%
\bibitem [{\citenamefont {Fioravanti}\ and\ \citenamefont {Gregori}(2021)}]{Fioravanti:2021dce}%
  \BibitemOpen
  \bibfield  {author} {\bibinfo {author} {\bibfnamefont {D.}~\bibnamefont {Fioravanti}}\ and\ \bibinfo {author} {\bibfnamefont {D.}~\bibnamefont {Gregori}},\ }\href@noop {} {\  (\bibinfo {year} {2021})},\ \Eprint {http://arxiv.org/abs/2112.11434} {arXiv:2112.11434 [hep-th]} \BibitemShut {NoStop}%
\bibitem [{\citenamefont {Dodelson}\ \emph {et~al.}(2023)\citenamefont {Dodelson}, \citenamefont {Grassi}, \citenamefont {Iossa}, \citenamefont {Panea~Lichtig},\ and\ \citenamefont {Zhiboedov}}]{Dodelson:2022yvn}%
  \BibitemOpen
  \bibfield  {author} {\bibinfo {author} {\bibfnamefont {M.}~\bibnamefont {Dodelson}}, \bibinfo {author} {\bibfnamefont {A.}~\bibnamefont {Grassi}}, \bibinfo {author} {\bibfnamefont {C.}~\bibnamefont {Iossa}}, \bibinfo {author} {\bibfnamefont {D.}~\bibnamefont {Panea~Lichtig}}, \ and\ \bibinfo {author} {\bibfnamefont {A.}~\bibnamefont {Zhiboedov}},\ }\href {\doibase 10.21468/SciPostPhys.14.5.116} {\bibfield  {journal} {\bibinfo  {journal} {SciPost Phys.}\ }\textbf {\bibinfo {volume} {14}},\ \bibinfo {pages} {116} (\bibinfo {year} {2023})},\ \Eprint {http://arxiv.org/abs/2206.07720} {arXiv:2206.07720 [hep-th]} \BibitemShut {NoStop}%
\bibitem [{\citenamefont {Jia}\ and\ \citenamefont {Rangamani}(2024)}]{Jia:2024zes}%
  \BibitemOpen
  \bibfield  {author} {\bibinfo {author} {\bibfnamefont {H.~F.}\ \bibnamefont {Jia}}\ and\ \bibinfo {author} {\bibfnamefont {M.}~\bibnamefont {Rangamani}},\ }\href@noop {} {\  (\bibinfo {year} {2024})},\ \Eprint {http://arxiv.org/abs/2408.05208} {arXiv:2408.05208 [hep-th]} \BibitemShut {NoStop}%
\bibitem [{\citenamefont {Arnaudo}\ \emph {et~al.}(2025{\natexlab{b}})\citenamefont {Arnaudo}, \citenamefont {Grassi},\ and\ \citenamefont {Hao}}]{Arnaudo:2025kof}%
  \BibitemOpen
  \bibfield  {author} {\bibinfo {author} {\bibfnamefont {P.}~\bibnamefont {Arnaudo}}, \bibinfo {author} {\bibfnamefont {A.}~\bibnamefont {Grassi}}, \ and\ \bibinfo {author} {\bibfnamefont {Q.}~\bibnamefont {Hao}},\ }\href@noop {} {\  (\bibinfo {year} {2025}{\natexlab{b}})},\ \Eprint {http://arxiv.org/abs/2502.01526} {arXiv:2502.01526 [hep-th]} \BibitemShut {NoStop}%
\bibitem [{\citenamefont {Litvinov}\ \emph {et~al.}(2014)\citenamefont {Litvinov}, \citenamefont {Lukyanov}, \citenamefont {Nekrasov},\ and\ \citenamefont {Zamolodchikov}}]{Litvinov:2013sxa}%
  \BibitemOpen
  \bibfield  {author} {\bibinfo {author} {\bibfnamefont {A.}~\bibnamefont {Litvinov}}, \bibinfo {author} {\bibfnamefont {S.}~\bibnamefont {Lukyanov}}, \bibinfo {author} {\bibfnamefont {N.}~\bibnamefont {Nekrasov}}, \ and\ \bibinfo {author} {\bibfnamefont {A.}~\bibnamefont {Zamolodchikov}},\ }\href {\doibase 10.1007/JHEP07(2014)144} {\bibfield  {journal} {\bibinfo  {journal} {JHEP}\ }\textbf {\bibinfo {volume} {07}},\ \bibinfo {pages} {144} (\bibinfo {year} {2014})},\ \Eprint {http://arxiv.org/abs/1309.4700} {arXiv:1309.4700 [hep-th]} \BibitemShut {NoStop}%
\bibitem [{\citenamefont {Piatek}\ and\ \citenamefont {Pietrykowski}(2019)}]{Piatek:2017fyn}%
  \BibitemOpen
  \bibfield  {author} {\bibinfo {author} {\bibfnamefont {M.}~\bibnamefont {Piatek}}\ and\ \bibinfo {author} {\bibfnamefont {A.~R.}\ \bibnamefont {Pietrykowski}},\ }\href {\doibase 10.1016/j.nuclphysb.2018.11.021} {\bibfield  {journal} {\bibinfo  {journal} {Nucl. Phys. B}\ }\textbf {\bibinfo {volume} {938}},\ \bibinfo {pages} {543} (\bibinfo {year} {2019})},\ \Eprint {http://arxiv.org/abs/1708.06135} {arXiv:1708.06135 [hep-th]} \BibitemShut {NoStop}%
\bibitem [{\citenamefont {Nekrasov}(2003)}]{Nekrasov:2002qd}%
  \BibitemOpen
  \bibfield  {author} {\bibinfo {author} {\bibfnamefont {N.~A.}\ \bibnamefont {Nekrasov}},\ }\href {\doibase 10.4310/ATMP.2003.v7.n5.a4} {\bibfield  {journal} {\bibinfo  {journal} {Adv. Theor. Math. Phys.}\ }\textbf {\bibinfo {volume} {7}},\ \bibinfo {pages} {831} (\bibinfo {year} {2003})},\ \Eprint {http://arxiv.org/abs/hep-th/0206161} {arXiv:hep-th/0206161} \BibitemShut {NoStop}%
\bibitem [{\citenamefont {Nekrasov}\ and\ \citenamefont {Okounkov}(2006)}]{Nekrasov:2003rj}%
  \BibitemOpen
  \bibfield  {author} {\bibinfo {author} {\bibfnamefont {N.}~\bibnamefont {Nekrasov}}\ and\ \bibinfo {author} {\bibfnamefont {A.}~\bibnamefont {Okounkov}},\ }\href {\doibase 10.1007/0-8176-4467-9_15} {\bibfield  {journal} {\bibinfo  {journal} {Prog. Math.}\ }\textbf {\bibinfo {volume} {244}},\ \bibinfo {pages} {525} (\bibinfo {year} {2006})},\ \Eprint {http://arxiv.org/abs/hep-th/0306238} {arXiv:hep-th/0306238} \BibitemShut {NoStop}%
\bibitem [{\citenamefont {Nekrasov}\ and\ \citenamefont {Shatashvili}(2010)}]{Nekrasov:2009rc}%
  \BibitemOpen
  \bibfield  {author} {\bibinfo {author} {\bibfnamefont {N.~A.}\ \bibnamefont {Nekrasov}}\ and\ \bibinfo {author} {\bibfnamefont {S.~L.}\ \bibnamefont {Shatashvili}},\ }in\ \href {\doibase 10.1142/9789814304634_0015} {\emph {\bibinfo {booktitle} {{16th International Congress on Mathematical Physics}}}}\ (\bibinfo {year} {2010})\ pp.\ \bibinfo {pages} {265--289},\ \Eprint {http://arxiv.org/abs/0908.4052} {arXiv:0908.4052 [hep-th]} \BibitemShut {NoStop}%
\bibitem [{\citenamefont {Bonelli}\ \emph {et~al.}(2023)\citenamefont {Bonelli}, \citenamefont {Iossa}, \citenamefont {Panea~Lichtig},\ and\ \citenamefont {Tanzini}}]{Bonelli:2022ten}%
  \BibitemOpen
  \bibfield  {author} {\bibinfo {author} {\bibfnamefont {G.}~\bibnamefont {Bonelli}}, \bibinfo {author} {\bibfnamefont {C.}~\bibnamefont {Iossa}}, \bibinfo {author} {\bibfnamefont {D.}~\bibnamefont {Panea~Lichtig}}, \ and\ \bibinfo {author} {\bibfnamefont {A.}~\bibnamefont {Tanzini}},\ }\href {\doibase 10.1007/s00220-022-04497-5} {\bibfield  {journal} {\bibinfo  {journal} {Commun. Math. Phys.}\ }\textbf {\bibinfo {volume} {397}},\ \bibinfo {pages} {635} (\bibinfo {year} {2023})},\ \Eprint {http://arxiv.org/abs/2201.04491} {arXiv:2201.04491 [hep-th]} \BibitemShut {NoStop}%
\bibitem [{\citenamefont {Lisovyy}\ and\ \citenamefont {Naidiuk}(2022)}]{Lisovyy2022}%
  \BibitemOpen
  \bibfield  {author} {\bibinfo {author} {\bibfnamefont {O.}~\bibnamefont {Lisovyy}}\ and\ \bibinfo {author} {\bibfnamefont {A.}~\bibnamefont {Naidiuk}},\ }\href {\doibase 10.1088/1751-8121/ac9ba7} {\bibfield  {journal} {\bibinfo  {journal} {Journal of Physics A: Mathematical and Theoretical}\ }\textbf {\bibinfo {volume} {55}},\ \bibinfo {pages} {434005} (\bibinfo {year} {2022})}\BibitemShut {NoStop}%
\bibitem [{\citenamefont {Besson}\ and\ \citenamefont {Jaramillo}(2025)}]{Besson:2024adi}%
  \BibitemOpen
  \bibfield  {author} {\bibinfo {author} {\bibfnamefont {J.}~\bibnamefont {Besson}}\ and\ \bibinfo {author} {\bibfnamefont {J.~L.}\ \bibnamefont {Jaramillo}},\ }\href {\doibase 10.1007/s10714-025-03438-6} {\bibfield  {journal} {\bibinfo  {journal} {Gen. Rel. Grav.}\ }\textbf {\bibinfo {volume} {57}},\ \bibinfo {pages} {110} (\bibinfo {year} {2025})},\ \Eprint {http://arxiv.org/abs/2412.02793} {arXiv:2412.02793 [gr-qc]} \BibitemShut {NoStop}%
\bibitem [{\citenamefont {Leaver}(1985)}]{Leaver:1985ax}%
  \BibitemOpen
  \bibfield  {author} {\bibinfo {author} {\bibfnamefont {E.~W.}\ \bibnamefont {Leaver}},\ }\href {\doibase 10.1098/rspa.1985.0119} {\bibfield  {journal} {\bibinfo  {journal} {Proc. Roy. Soc. Lond. A}\ }\textbf {\bibinfo {volume} {402}},\ \bibinfo {pages} {285} (\bibinfo {year} {1985})}\BibitemShut {NoStop}%
\bibitem [{\citenamefont {Matone}(1995)}]{Matone:1995rx}%
  \BibitemOpen
  \bibfield  {author} {\bibinfo {author} {\bibfnamefont {M.}~\bibnamefont {Matone}},\ }\href {\doibase 10.1016/0370-2693(95)00920-G} {\bibfield  {journal} {\bibinfo  {journal} {Phys. Lett. B}\ }\textbf {\bibinfo {volume} {357}},\ \bibinfo {pages} {342} (\bibinfo {year} {1995})},\ \Eprint {http://arxiv.org/abs/hep-th/9506102} {arXiv:hep-th/9506102} \BibitemShut {NoStop}%
\end{thebibliography}%

\clearpage

\onecolumngrid

\section*{SUPPLEMENTAL MATERIAL}
\subsection{Longer formulas}

The roots $R_{\pm}$ of the equation $r\,f(r)=0$ are expressed in terms of $R_h$ and $\Lambda$ by 
\begin{equation}
R_{\pm}=\frac{-R_h\pm\sqrt{\frac{12}{\Lambda}-3 R_h^2}}{2}.
\end{equation}

The dictionary for the differential equation \eqref{heunnormalform} is
\begin{equation}\label{dictioheun}
\begin{aligned}
x&=\frac{R_h (R_+-R_-)}{R_+ (R_h-R_-)},\\
a_0&=\frac{3 i R_- \omega }{\Lambda  (R_--R_h) (R_--R_+)},\\
a_1&=\frac{3 i R_h \omega }{\Lambda  (R_h-R_-) (R_+-R_h)},\\
a_x&=\frac{3 i R_+ \omega }{\Lambda  (R_+-R_h) (R_+-R_-)},\\
a_{\infty}&=s,\\
u&=\frac{3 \ell(\ell+1)}{\Lambda  \left(R_h^2-R_+^2\right)}+\frac{18 R_+^2 \omega ^2 \left(-2 R_h^2-2 R_h R_++R_+^2\right)}{\Lambda ^2 (R_h-R_+)^3 (R_h+R_+) (R_h+2 R_+)^2}+\frac{3-3 s^2}{\Lambda  \left(R_h^2-R_+^2\right)}-\frac{2 R_h^2+2 R_h R_+-2 R_+^2 s^2+R_+^2}{2 R_h^2-2 R_+^2}.
\end{aligned}
\end{equation}
The wave function redefinition in \eqref{wfredef} requires the function
\begin{equation}\label{pz}
p(z)=\frac{1}{\sqrt{z(1-z)} \sqrt{R_h (R_-+R_+ (z-1))-R_- R_+ z}}.
\end{equation}

In terms of the parameters
\begin{equation}
\begin{aligned}
\alpha&=1-a_0-a_1-a_x+a_{\infty},\\
\beta&=1-a_0-a_1-a_x-a_{\infty},\\
\gamma&=1-2a_0,\\
\delta&=1-2a_1,\\
\epsilon&=1-2a_x,\\
q&=\frac{1}{2}+x\left(a_0^2+a_1^2+a_x^2-a_{\infty}^2\right)-a_x-a_1\,x+a_0\left[2a_x-1+x\left(2a_1-1\right)\right]+\left(1-x\right)u,
\end{aligned}
\end{equation}
the local solution satisfying ingoing boundary condition at the horizon is
\begin{equation}\label{insol}
\begin{aligned}
\psi_{\text{in}}(z)&=z^{\gamma /2} (1-z)^{\delta /2} (z-x)^{\epsilon /2}(1-x)^{-\epsilon /2}\,\text{Heun}\left(1-x,\alpha  \beta -q,\alpha ,\beta ,\delta ,\gamma ,1-z\right),
\end{aligned}
\end{equation}
and the one satisfying outgoing boundary condition at the cosmological horizon is
\begin{equation}\label{upsol}
\begin{aligned}
\psi_{\text{up}}(z)&=z^{\gamma /2} (1-z)^{\delta /2}x^{-\gamma /2} (1-x)^{-\delta /2} (z-x)^{\epsilon /2}\,\text{Heun}\left(\frac{x}{x-1},\frac{q-\alpha  \beta  x}{1-x},\alpha ,\beta ,\epsilon ,\delta ,\frac{z-x}{1-x}\right).
\end{aligned}
\end{equation}
The discarded local solutions (with respect to the QNMs boundary conditions) are the local solution satisfying outgoing boundary condition at the horizon:
\begin{equation}\label{psiout}
\begin{aligned}
\psi_{\text{out}}(z)=&z^{\gamma /2} (1-z)^{1-\delta /2} (z-x)^{\epsilon /2}(1-x)^{-\epsilon /2}\\
&\times\text{Heun}\Bigl(1-x,[(1-x)\gamma+\epsilon](1-\delta)+\alpha  \beta -q,\alpha+1-\delta,\beta+1-\delta ,2-\delta ,\gamma ,1-z\Bigr),
\end{aligned}
\end{equation}
and the local solution satisfying ingoing boundary condition at the cosmological horizon:
\begin{equation}\label{psidown}
\begin{aligned}
&\psi_{\text{down}}(z)=z^{\gamma /2} (1-z)^{\delta /2}x^{-\gamma /2} (1-x)^{-\delta /2} (z-x)^{1-\epsilon /2}\\
&\times\text{Heun}\biggl(\frac{x}{x-1},\frac{-q+x\,(\beta-\gamma-\delta)\,(\alpha-\gamma-\delta)+\gamma\,(\alpha+\beta-\gamma-\delta)}{x-1},-\alpha+\gamma+\delta,-\beta+\gamma+\delta,2-\epsilon,\delta,\frac{z-x}{1-x}\biggr).
\end{aligned}
\end{equation}

We define the tortoise coordinate in the Schwarzschild-de Sitter geometry by
\begin{equation}\label{tortoise}
\begin{aligned}
r_*=\,&-\frac{3 \left[R_h (R_--R_+) \log \left(\frac{r}{R_h}-1\right)+R_- (R_+-R_h) \log \left(\frac{r}{R_h}-\frac{R_-}{R_h}\right)+R_+ (R_h-R_-) \log \left(\frac{R_+}{R_h}-\frac{r}{R_h}\right)\right]}{\Lambda  (R_h-R_-) (R_h-R_+) (R_--R_+)}\\
&+\frac{1}{2} R_h \left[\log\left(\frac{3}{\Lambda\,R_h^2}\right)-1\right],
\end{aligned}
\end{equation}
which, in the $\Lambda\to 0^+$ limit, reduces to the Schwarzschild tortoise coordinate
\begin{equation}
r_*^{(\text{Schw})}=r+R_h\,\log \left(\frac{r}{R_h}-1\right).
\end{equation}

The connection coefficients for the Heun equation have the following expressions
\begin{equation}\label{Cupin}
\begin{aligned}
C_{\text{up},\text{in}} &=\sum_{\sigma=\pm}\frac{\Gamma\left(1-2a_{x}\right)\Gamma\left(-2\sigma a\right)\Gamma\left(1-2\sigma a\right)\Gamma\left(2a_{1}\right)}{\prod_{\pm}\Gamma\left(\frac{1}{2}- a_{x}-\sigma a\pm a_0\right)\Gamma\left(\frac{1}{2}-\sigma a+a_{1}\pm a_\infty\right)}x^{-a_x+\sigma a}e^{-\frac{\sigma}{2}\partial_{a}F(x)-\frac{1}{2}\partial_{a_{x}}F(x)+\frac{1}{2}\partial_{a_{1}}F(x)},
\end{aligned}
\end{equation}
\begin{equation}\label{Cupout}
\begin{aligned}
C_{\text{up},\text{out}} &=\sum_{\sigma=\pm}\frac{\Gamma\left(1-2a_{x}\right)\Gamma\left(-2\sigma a\right)\Gamma\left(1-2\sigma a\right)\Gamma\left(-2a_{1}\right)}{\prod_{\pm}\Gamma\left(\frac{1}{2}- a_{x}-\sigma a\pm a_0\right)\Gamma\left(\frac{1}{2}-\sigma a-a_{1}\pm a_\infty\right)}x^{-a_x+\sigma a}e^{-\frac{\sigma}{2}\partial_{a}F(x)-\frac{1}{2}\partial_{a_{x}}F(x)-\frac{1}{2}\partial_{a_{1}}F(x)},
\end{aligned}
\end{equation}
\begin{equation}\label{Cinup}
\begin{aligned}
C_{\text{in},\text{up}} &=e^{-\frac{1}{2}\partial_{a_{1}}F(x)}\sum_{\sigma=\pm}\frac{\Gamma\left(1-2a_{1}\right)\Gamma\left(-2\sigma a\right)\Gamma\left(1-2\sigma a\right)\Gamma\left(2a_{x}\right)}{\prod_{\pm}\Gamma\left(\frac{1}{2}- a_{1}-\sigma a\pm a_\infty\right)\Gamma\left(\frac{1}{2}-\sigma a+a_{x}\pm a_0\right)}x^{\sigma a}e^{-\frac{\sigma}{2}\partial_{a}F(x)}x^{a_{x}}e^{\frac{1}{2}\partial_{a_{x}}F(x)}.
\end{aligned}
\end{equation}
\begin{equation}\label{Cindown}
\begin{aligned}
C_{\text{in},\text{down}} &=e^{-\frac{1}{2}\partial_{a_{1}}F(x)}\sum_{\sigma=\pm}\frac{\Gamma\left(1-2a_{1}\right)\Gamma\left(-2\sigma a\right)\Gamma\left(1-2\sigma a\right)\Gamma\left(-2a_{x}\right)}{\prod_{\pm}\Gamma\left(\frac{1}{2}- a_{1}-\sigma a\pm a_\infty\right)\Gamma\left(\frac{1}{2}-\sigma a-a_{x}\pm a_0\right)}x^{\sigma a}e^{-\frac{\sigma}{2}\partial_{a}F(x)}x^{-a_{x}}e^{-\frac{1}{2}\partial_{a_{x}}F(x)}.
\end{aligned}
\end{equation}

The ratio of connection coefficients used in the analytic computations can be written as 
\begin{equation}\label{ratioconncoeff}
\begin{aligned}
\frac{C_{\text{up},\text{in}}}{C_{\text{up},\text{out}}}=e^{\partial_{a_1} F(x)}\frac{\mathcal{C}^{\text{exact}}_{a_1,a}-\mathcal{G}^{\text{exact}}\,\mathcal{C}^{\text{exact}}_{a_1,-a}}{\mathcal{C}^{\text{exact}}_{-a_1,a}-\mathcal{G}^{\text{exact}}\,
\mathcal{C}^{\text{exact}}_{-a_1,-a}},
\end{aligned}
\end{equation}
with
\begin{align}
\mathcal{C}^{\text{exact}}_{a_1,a}&=\frac{\Gamma(2a_1)}{\prod_{\pm}\Gamma\left(\frac{1}{2}+a_1+a\pm a_{\infty}\right)},\label{greensparts1}\\
\mathcal{G}^{\text{exact}}&=x^{2a}\,e^{-\partial_aF(x)}\,\frac{\Gamma(-2a)^2\prod_{\pm}\Gamma\left(\frac{1}{2}-a_{x}+a\pm a_0\right)}{\Gamma(2a)^2\prod_{\pm}\Gamma\left(\frac{1}{2}-a_{x}-a\pm a_0\right)}.\label{greensparts2}
\end{align}

The parameter $a$ is defined as an instanton expansion
\begin{equation}
a=\sum_{k\ge 0}a^{(k)}_{\text{inst}}\,x^k,
\end{equation}
as explained in the next Appendix. The zeroth instanton order of $a$ reads
\begin{equation}\label{a0inst}
a^{(0)}_{\text{inst}}=\frac{1}{2} \sqrt{1+4 \ell(\ell+1)-12 R_h^2 \omega ^2}+\mathcal{O}\left(\sqrt{\Lambda}\right),
\end{equation}
in the small $\Lambda$ expansion. By adding more instanton contributions, every order in the small $\Lambda$ expansion gets modified if $\omega$ is assumed to be of order $\Lambda^0$. Instead, if the frequency is assumed to be of order $\sqrt{\Lambda}$, in terms of the rescaled frequency $\tilde{\omega}\equiv \omega/\sqrt{\Lambda}$, the $\Lambda$-expansion is consistent with the instanton expansion. In particular, the first order correction of the parameter $a$ in the small $\Lambda$ expansion reads\footnote{This result is obtained by using the second order in the instanton expansion.}
\begin{equation}\label{a1}
\begin{aligned}
a_{\Lambda}^{(1)}=\frac{R_h^2 \left(L \left(15 L^2-6 L \left(s^2+1\right)-s^4+5 s^2-4\right)-3 \tilde{\omega}^2 \left(6 (L-1) s^2+L (15 L-11)+3 s^4\right)\right)}{6 (2 \ell+1) L (4 L-3)},
\end{aligned}
\end{equation}
where $L=\ell(\ell+1)$.

By considering more instanton contributions, it is also possible to find the subleading orders in the frequencies \eqref{dSmodes}, from the poles of the $\Gamma$ function $\Gamma\left(\frac{1}{2}+a-a_x+a_0\right)$. For instance, with the first instanton contribution, the frequencies read
\begin{equation}
\omega_n^{(\text{dS})}=-i\,\sqrt{\frac{\Lambda}{3}}\left(\ell+n+1\right)+\omega_1\,\Lambda^{3/2}+\mathcal{O}\left(\Lambda^2\right),
\end{equation}
with
\begin{equation}
\begin{aligned}
\omega_1=&-\frac{i R_h^2}{24 \sqrt{3} \ell (\ell+1) (2 \ell+1) (4 \ell (\ell+1)-3)} \biggl[\ell (\ell+1) \left(\ell^2 (60 n (n+1)+22)+\ell \left(60 n^2+62 n+23\right)-n (44 n+43)-15\right)\\
&+4 s^4 \left(2 \ell^2+\ell (6 n+5)+3 (n+1)^2\right)+4 s^2 \left(6 \ell^3 (2 n+1)+\ell^2 (6 n (n+4)+11)+\ell \left(6 n^2-1\right)-6 (n+1)^2\right)\biggr].
\end{aligned}
\end{equation}

The small $x$ expansion of the ratio of the connection coefficients $C_{\text{up},\text{in}}/C_{\text{up},\text{out}}$ reads
\begin{equation}\label{ratioexp}
\begin{aligned}
&\frac{C_{\text{up},\text{in}} }{C_{\text{up},\text{out}}}=\frac{\Gamma\left(2 a_{1}\right)}{\Gamma\left(-2 a_{1}\right)}\prod_{\pm}\frac{\Gamma\left(\frac{1}{2}+ a- a_{1}\pm a_\infty\right)}{\Gamma\left(\frac{1}{2}+a+a_{1}\pm a_\infty\right)}+\\
&x^{2a}\,
\,\biggl[\frac{\Gamma\left(-2a\right)\Gamma\left(1-2a\right)\Gamma\left(2a_1\right)\prod_{\pm}\Gamma\left(\frac{1}{2}-a_{x}+a\pm a_0\right)\Gamma\left(\frac{1}{2}-a_{1}+a\pm a_\infty\right)}{\Gamma\left(2a\right)\Gamma\left(1+2a\right)\Gamma\left(-2a_1\right)\prod_{\pm}\Gamma\left(\frac{1}{2}-a_{x}-a\pm a_0\right)\Gamma\left(\frac{1}{2}+a_{1}-a\pm a_\infty\right)}\\
&-\frac{\Gamma\left(-2a\right)\Gamma\left(1-2a\right)\Gamma\left(2a_{1}\right)\prod_{\pm}\Gamma\left(\frac{1}{2}- a_{x}+a\pm a_0\right)\Gamma\left(\frac{1}{2}+a-a_{1}\pm a_\infty\right)^2}{\Gamma\left(2a\right)\Gamma\left(1+2a\right)\Gamma\left(-2a_{1}\right)\prod_{\pm}\Gamma\left(\frac{1}{2}+ a+a_{1}\pm a_\infty\right)\Gamma\left(\frac{1}{2}- a_{x}-a\pm a_0\right)\Gamma\left(\frac{1}{2}-a-a_{1}\pm a_\infty\right)}\biggr]+\mathcal{O}\left(x^{2a+1}\right).
\end{aligned}
\end{equation}

The $\rho_h$ factor in \eqref{replacement} is given by
\begin{equation}\label{rhoh}
\begin{aligned}
\rho_h=&\left(\frac{e \Lambda }{3}\right)^{\frac{i R_h \omega }{2}} R_h^{i R_h \omega  \left(1+\frac{3}{\Lambda  (R_h-R_-) (R_h-R_+)}\right)} (R_h-R_-)^{\frac{3 i R_h \omega }{\Lambda  (R_h-R_-) (R_h-R_+)}+\frac{3 i R_- \omega }{\Lambda  (R_--R_h) (R_--R_+)}}\\
&\times(-R_-)^{\frac{1}{2}-\frac{3 i R_h \omega }{\Lambda  (R_h-R_-) (R_h-R_+)}} (R_+-R_h)^{\frac{1}{2}+\frac{3 i R_+ \omega }{\Lambda  (R_+-R_h) (R_+-R_-)}}.
\end{aligned}
\end{equation}

The function $\mathfrak{G}$ appearing in the residue sum \eqref{residuesum} is given by
\begin{equation}\label{frakG}
\begin{aligned}
\mathfrak{G}&=\frac{\Gamma (2 a_1) \Gamma (-2 a_0-n) \Gamma (2 a_0-2 a_x+2 n+2) \Gamma (2 a_0-2 a_x+2 n+1)\prod_{\pm} \Gamma (-a_0-a_1\pm a_\infty+a_x-n)}{\Gamma (-2 a_1) \Gamma (-2 a_x+n+1) \Gamma (-2 a_0+2 a_x-2 n-1) \Gamma (-2 a_0+2 a_x-2 n) \Gamma (2 a_0-2 a_x+n+1) }\\
&\times\left[\prod_{\pm}\frac{1}{\Gamma (a_0+a_1\pm a_\infty-a_x+n+1)}-\prod_{\pm}\frac{\Gamma (-a_0-a_1\pm a_\infty+a_x-n)}{\Gamma (-a_0+a_1\pm a_\infty+a_x-n)  \Gamma (a_0-a_1\pm a_\infty-a_x+n+1) }\right].
\end{aligned}
\end{equation}

The following expansion is useful for the terms appearing in the residue sum \eqref{residuesum}:
\begin{equation}\label{expansioncoeff}
\begin{aligned}
&\frac{3 R_h R_-}{\Lambda\,(R_h-R_-)\,2a_1\,\rho_h^2}\mathfrak{G}\,\left(\frac{d(a+a_0-a_x)}{d \omega }\right)^{-1}\bigg|_{\omega=\omega_n^{\text{dS}}}=\\
&\frac{(-1)^{\ell+n-\frac{1}{2}}\, (\ell+n+1)\,  (2 \ell+n+1)!\,(\ell!)^2 \,[(\ell-s)!]^2\,[(\ell+s)!]^2}{3 \, [(2 \ell)!]^2\, [(2 \ell+1)!]^2}\,R_h^2\,\Lambda+\mathcal{O}\left(\Lambda^{2}\right).
\end{aligned}
\end{equation}

The coefficient $C$ is given by 
\be\label{Cdef}
C =   \frac{(-1)^{\ell}\,2^{2\ell+2}\,(\ell+1)\,(\ell!)^2 \,[(\ell-s)!]^2\,[(\ell+s)!]^2}{ [(2 \ell)!]^2\,(2 \ell+1)!}.
\ee


Taking the expansion of $\frac{C_{\text{up},\text{in}}}{C_{\text{up},\text{out}}}$ at zero instantons using \eqref{a0inst}, and replacing the term $x^{2a}\frac{\Gamma\left(\frac{1}{2}-a_x+a_0+a\right)}{\Gamma\left(\frac{1}{2}-a_x+a_0-a\right)}$ in $\mathcal{G}^{\text{exact}}$ with $\left(-2 i \omega R_h\right)^{2a^{(0)}_{\text{inst}}}$ from \eqref{confluencecut}, 
we get
\begin{equation}\label{ratioconncoeffcut}
\begin{aligned}
&\frac{\tilde{C}_{\text{up},\text{in}}}{\tilde{C}_{\text{up},\text{out}}}\sim\frac{\mathcal{C}^{\text{exact}}_{a_1,a^{(0)}_{\text{inst}}}-\frac{\Gamma\left(-2a^{(0)}_{\text{inst}}\right)^2}{\Gamma\left(2a^{(0)}_{\text{inst}}\right)^2}\,\frac{\Gamma\left(\frac{1}{2}-a_x+a^{(0)}_{\text{inst}}-a_0\right)}{\Gamma\left(\frac{1}{2}-a_x-a^{(0)}_{\text{inst}}-a_0\right)}\,\mathcal{C}^{\text{exact}}_{a_1,-a^{(0)}_{\text{inst}}}\left(-2 i \omega R_h\right)^{2a^{(0)}_{\text{inst}}}}{\mathcal{C}^{\text{exact}}_{-a_1,a^{(0)}_{\text{inst}}}-\frac{\Gamma\left(-2a^{(0)}_{\text{inst}}\right)^2}{\Gamma\left(2a^{(0)}_{\text{inst}}\right)^2}\,\frac{\Gamma\left(\frac{1}{2}-a_x+a^{(0)}_{\text{inst}}-a_0\right)}{\Gamma\left(\frac{1}{2}-a_x-a^{(0)}_{\text{inst}}-a_0\right)}\,
\mathcal{C}^{\text{exact}}_{-a_1,-a^{(0)}_{\text{inst}}}\left(-2 i \omega R_h\right)^{2a^{(0)}_{\text{inst}}}}=\frac{\Gamma (2 a_1) \prod_{\pm}\Gamma \left(\frac{1}{2}+a^{(0)}_{\text{inst}}-a_1\pm a_{\infty}\right)}{\Gamma (-2 a_1) \prod_{\pm}\Gamma \left(\frac{1}{2}+a^{(0)}_{\text{inst}}+a_1\pm a_{\infty}\right)}\\
&+\frac{\Gamma (-2 a^{(0)}_{\text{inst}})^2 \Gamma (2 a_1) \Gamma \left(\frac{1}{2}+a^{(0)}_{\text{inst}}-a_0-a_x\right) \prod_{\pm}\Gamma \left(\frac{1}{2}+a^{(0)}_{\text{inst}}-a_1\pm a_{\infty}\right)}{\Gamma (2 a^{(0)}_{\text{inst}})^2 \Gamma (-2 a_1) \Gamma \left(\frac{1}{2}-a^{(0)}_{\text{inst}}-a_0-a_x\right)}\Biggl[-\frac{1}{\prod_{\pm}\Gamma \left(\frac{1}{2}-a^{(0)}_{\text{inst}}+a_1\pm a_{\infty}\right)}\\
&+\frac{\prod_{\pm}\Gamma \left(\frac{1}{2}+a^{(0)}_{\text{inst}}-a_1\pm a_{\infty}\right)}{\prod_{\pm}\Gamma \left(\frac{1}{2}-a^{(0)}_{\text{inst}}-a_1\pm a_{\infty}\right)\prod_{\pm}\Gamma \left(\frac{1}{2}+a^{(0)}_{\text{inst}}+a_1\pm a_{\infty}\right)}\Biggr]\left(-2 i \omega R_h\right)^{2a^{(0)}_{\text{inst}}}+\mathcal{O}\left[\left(-2 i \omega R_h\right)^{4a^{(0)}_{\text{inst}}}\right].
\end{aligned}
\end{equation}

The branch cut structure in $\omega$ is then explicit by looking at the expansion
\begin{equation}\label{logcut}
\left(-2 i \omega R_h\right)^{2a^{(0)}_{\text{inst}}}=\left(-2 i \omega R_h\right)^{2\ell+1}+\frac{3\,(-2 i R_h \omega )^{2 \ell+3}}{2 (2 \ell+1)}\log\left(-2 i\omega R_h\right)+\mathcal{O}\left(\omega^{2\ell+5}\right).
\end{equation}

The coefficient of $\left(-2 i \omega R_h\right)^{2a^{(0)}_{\text{inst}}}$ in \eqref{ratioconncoeffcut} admits the expansion
\begin{equation}\label{coeffcutomegaexp1}
\begin{aligned}
&\frac{\Gamma (-2 a^{(0)}_{\text{inst}})^2 \Gamma (2 a_1) \Gamma \left(\frac{1}{2}+a^{(0)}_{\text{inst}}-a_0-a_x\right) \prod_{\pm}\Gamma \left(\frac{1}{2}+a^{(0)}_{\text{inst}}-a_1\pm a_{\infty}\right)}{\Gamma (2 a^{(0)}_{\text{inst}})^2 \Gamma (-2 a_1) \Gamma \left(\frac{1}{2}-a^{(0)}_{\text{inst}}-a_0-a_x\right)}\\
&\times\left[\frac{\prod_{\pm}\Gamma \left(\frac{1}{2}+a^{(0)}_{\text{inst}}-a_1\pm a_{\infty}\right)}{\prod_{\pm}\Gamma \left(\frac{1}{2}-a^{(0)}_{\text{inst}}-a_1\pm a_{\infty}\right)\prod_{\pm}\Gamma \left(\frac{1}{2}+a^{(0)}_{\text{inst}}+a_1\pm a_{\infty}\right)}-\frac{1}{\prod_{\pm}\Gamma \left(\frac{1}{2}-a^{(0)}_{\text{inst}}+a_1\pm a_{\infty}\right)}\right]=\\
&\frac{\pi  (-1)^{\ell} (2 \ell+1) ((\ell-s)!)^2 ((\ell+s)!)^2}{3\ 4^{2 \ell} \Gamma \left(\ell+\frac{1}{2}\right)^2 \Gamma (2 \ell+2)^2}+\mathcal{O}\left(\Lambda,\omega\right).
\end{aligned}
\end{equation}
The following expansion is also useful for the computation of the integral discontinuity:
\begin{equation}\label{coeffcutomegaexp2}
\begin{aligned}
\frac{-3 R_h R_-}{\Lambda\,(R_h-R_-)}\,\frac{1}{2a_1\,\rho_h^2}=-\frac{i}{2 \omega }+\mathcal{O}\left(\sqrt{\Lambda},\omega^0\right).
\end{aligned}
\end{equation}

\subsection{The gauge theory framework and conventions}

In this section, we introduce the main contributions useful for defining the instanton partition function appearing in the connection formulas and we introduce the conventions used. We denote with $\vec{Y}=(Y_1,Y_2)$ a pair of Young diagrams, with $\vec{a}=(a_1,a_2)$ the vacuum expectation value of the scalar in the vector multiplet, and with $\epsilon_1,\epsilon_2$ the parameters characterizing the $\Omega$-background. We define the hypermultiplet and vector contributions as
\begin{equation}
\begin{aligned}
z_{\text{hyp}} \left( \vec{a}, \vec{Y}, m \right)&=\prod_{k= 1,2} \prod_{(i,j) \in Y_k} \left[ a_k + m + \epsilon_1 \left( i - \frac{1}{2} \right) + \epsilon_2 \left( j - \frac{1}{2} \right) \right],\\
z_{\text{vec}} \left( \vec{a}, \vec{Y} \right)&=\prod_{i,j=1}^2\prod_{s\in Y_i}\frac{1}{a_i-a_j-\epsilon_1L_{Y_j}(s)+\epsilon_2(A_{Y_i}(s)+1)}\prod_{t\in Y_j}\frac{1}{-a_j+a_i+\epsilon_1(L_{Y_i}(t)+1)-\epsilon_2A_{Y_j}(s)}.
\end{aligned}
\end{equation}
We always adopt the conventions $\epsilon_1=1$ and $\vec{a}=(a,-a)$.
By denoting with $m_1,m_2,m_3,m_4$ the masses of the four fundamental hypermultiplets, these are defined in terms of the monodromy parameters $a_0,a_x,a_1,a_{\infty}$ as
\begin{equation}\label{gaugemasses}
\begin{aligned}
m_1&=-a_x-a_0,\quad
&m_2=-a_x+a_0,\\
m_3&=a_{\infty}+a_1,\quad
&m_4=-a_{\infty}+a_1.
\end{aligned}
\end{equation}
The position of the fourth singularity of the Heun equation, denoted with $x$, corresponds to the instanton counting parameter $x=e^{2\pi i\tau}$, where $\tau$ is related to the gauge coupling $g_{\rm YM}$ by 
\begin{equation}
\tau=\frac{\theta}{2\pi}+i\frac{4\pi}{g_{\rm YM}^2}.
\end{equation}
The instanton part of the NS free energy is given as a power series in $x$ by
\begin{equation}
F(x)=\lim_{\epsilon_2\to 0}\epsilon_2\log\Biggl[(1-x)^{-2\epsilon_2^{-1}\left(\frac{1}{2}+a_1\right)\left(\frac{1}{2}+a_x\right)}\sum_{\vec{Y}}x^{|\vec{Y}|}z_{\text{vec}} \left( \vec{a}, \vec{Y} \right)\prod_{i=1}^4z_{\text{hyp}} \left( \vec{a}, \vec{Y}, m_i \right)\Biggr],
\end{equation}
which can be expanded as
\begin{equation}
F(x)=\frac{\left(4 a^2-4 a_0^2+4 a_x^2-1\right) \left(4 a^2+4 a_1^2-4 a_\infty^2-1\right)}{8-32 a^2}\,x+\mathcal{O}\left(x^2\right).
\end{equation}

The gauge parameter $a$ parametrizes the composite monodromy around the points $z=0$ and $z=x$, and is expressed as a series expansion in the instanton counting parameter $x$, obtained by inverting the Matone relation \cite{Matone:1995rx}
\begin{equation}
u =-\frac{1}{4} - a^2 + a_x^2 + a_0^2 + x\, \partial_x F(x),
\end{equation}
where the parameter $u$, appearing in the differential equation \eqref{heunnormalform} as the accessory parameter, corresponds to the complex modulus parametrizing the energy of the corresponding Seiberg-Witten curve.
The instanton expansion of $a$ reads 
\begin{equation}\label{agauge}
a=\pm\left\{\sqrt{-\frac{1}{4}-u+a_x^2+a_0^2}+\frac{\bigl(\frac{1}{2}+u-a_x^2-a_0^2-a_1^2+a_{\infty}^2\Bigr)\Bigl(\frac{1}{2}+u-2a_x^2\Bigr)}{2(1+2u-2a_x^2-2a_0^2)\sqrt{-\frac{1}{4}-u+a_x^2+a_0^2}}\,x+\mathcal{O}(x^2)\right\}.
\end{equation}
As can be seen, $a$ is in principle defined up to a sign. We always choose the branch for which $a$ has a positive real part.

In the analytic computations, we work at zero instantons. This means that the expansion of $a$ is truncated at the first term in \eqref{agauge}, $a=\pm \sqrt{-\frac{1}{4}-u+a_x^2+a_0^2}$\,, and $F(x)=0$.

\end{document}